\documentclass[runningheads]{llncs}
\usepackage{graphicx}
\usepackage{color}
\usepackage{booktabs, threeparttable}
\usepackage{array}
\newcolumntype{L}[1]{>{\raggedright\arraybackslash}p{#1}}
\usepackage{url}
\usepackage{todonotes}
\usepackage{siunitx}
\makeatletter
\newcommand*\titleheader[1]{\gdef\@titleheader{#1}}
\AtBeginDocument{%
  \let\st@red@title\@title
  \def\@title{%
    \bgroup\normalfont\large\centering\@titleheader\par\egroup
    \vskip1.5em\st@red@title}
}
\makeatother

\usepackage{fancyhdr}
\begin{document}
%Set page style.

\thispagestyle{fancy} 
%\history{Date of publication xxxx 00, 0000, date of current version xxxx 00, 0000.}
%
%\title{\textcolor{red}{On specifics of NR-based directional access \\for system-level evaluation?}
\title{A Concise Review of 5G New Radio Capabilities for Directional Access at mmWave Frequencies}
\titleheader{Date of publication }
%\thanks{Supported by organization x.}
%
\titlerunning{A Concise Review of 5G New Radio Capabilities}

% If the paper title is too long for the running head, you can set
% an abbreviated paper title here
%
\author{Giulia Sanfilippo\inst{1,2} \and
Olga Galinina\inst{3,4}\orcidID{0000-0002-5386-1061}\and\\
Sergey Andreev\inst{3}\orcidID{0000-0003-4049-8715}
\and 
Sara Pizzi\inst{1}\orcidID{0000-0002-0550-5978}\and\\
Giuseppe Araniti\inst{1}\orcidID{0000-0001-8670-9413}}
\authorrunning{G. Sanfilippo et al.}
% First names are abbreviated in the running head.
% If there are more than two authors, 'et al.' is used.
%
\institute{Mediterranean University of Reggio Calabria, Italy\\ \and 
Vodafone Italy, Italy\\
\and Tampere University of Technology, Finland\\
\email{giuliasnfp@gmail.com, \{olga.galinina,sergey.andreev\}@tut.fi,\{sara.pizzi,araniti\}@unirc.it}\\
%\url{http://www.springer.com/gp/computer-science/lncs}
 }
\maketitle              % typeset the header of the contribution

\setcounter{footnote}{0}

\vspace{-5px}
\begin{abstract}
In this work, we briefly outline the core 5G air interface improvements introduced by the latest New Radio (NR) specifications, as well as elaborate on the unique features of initial access in 5G NR with a particular emphasis on millimeter-wave (mmWave) frequency range. The highly directional nature of 5G mmWave cellular systems poses a variety of fundamental differences and research problem formulations, and a holistic understanding of the key system design principles behind the 5G NR is essential. Here, we condense the relevant information collected from a wide diversity of 5G NR standardization documents (based on 3GPP Release 15) to distill the essentials of directional access in 5G mmWave cellular, which becomes the foundation for any corresponding system-level analysis. 
%Using our reference scenario as a typical example, we analyze selected performance metrics and provide insights into challenges of beam-forming and directional access in general.
\keywords{mmWave \and beamforming \and New Radio \and 5G NR \and numerology\and initial access\and random access.}
\end{abstract}\vspace{-5px}
%
%
%

%docomo2018,Yifei2017,
%M. Giordani, M. Mezzavilla, M. Zorzi, “Initial Access in 5G mmWave Cellular Networks”, IEEE Communication
%Magazine, vol.54, Issue 11, pp 40-47, November 2016.
%• H. Shokri-Ghadikolaei, C. Fischione, G. Fodor, P. Popovski, M.Zorzi, “Millimeter Wave Cellular Networks: A MAC Layer
%Perspective” IEEE Transaction on Communications, vol.63, NO.10, pp 3437-3458, October 2015.
%• S.Y. Lien, S.L. Shieh, Y. Huang, B. Su, Y.L. Hsu, H.Y. Wei, “5G New Radio: Waveform, Frame Structure, Multiple Access
%and Initial Access”, IEEE Communication Magazine, vol.55, Issue 6, pp 64-71, June 2017.

%----------------------------------------------------------------------------------------------------------------------------------------------------------------------------------------------------------------------------------------------------------------------------------%

\vspace{-10px}
\section{Introduction}
\vspace{-5px}

%\cite{mediatek20185Gdesign}

In December 2017, the Third Generation Partnership Project (3GPP) released an early version of the first 5G specifications \cite{release15} -- officially named 5G \textit{non-standalone} (NSA) -- to enable 5G New Radio (NR) deployments on top of the current 4G systems. In this case, a device fully relies on the existing LTE interface and protocols for control procedures while the data traffic can be split between the 5G NR and LTE, which corresponds to architecture option three: ``LTE assisted, EPC\footnote{Evolved Packet Core} Connected"~\cite{RP161266}. % (LTE EUTRAN and EPC). 

After this initial phase, further developments set the course for 5G \textit{standalone} (SA) operation by incorporating a complete set of specifications for the new 5G Core Network complementing the NSA version to enable operation not relying on the 4G infrastructure. Half a year later, in June 2018, the complete SA description has been ``frozen" in Release 15. This signifies that its technical specifications\footnote{3GPP naming convention: TS = technical specifications, TR = technical report, CR = correction request.} are considered sufficiently stable, i.e., all new features, along with the functionality required to implement them have been defined and addressed in the standardization documentation. 
%that it may be considered "frozen". That is, only CRs for necessary corrections of errors shall be considered.   incorporated

The completion of the SA 5G NR specifications not only opens the door to deploying 5G networks without relying on the existing infrastructure but also marks a decisive step into a new era of an interconnected society. 
%"but also brings a brand new end-to-end network architecture, making 5G a facilitator and an accelerator during the intelligent information and communications technology improvement process of enterprise customers and vertical industries." 
Aiming at aggressive performance targets, the ongoing 3GPP efforts revolve around the following three emerging use-cases: (i) enhanced mobile broadband (eMBB) with the data rate requirements of up to 10 to 20 Gbps and support for high mobility (up to \si{500}{ km/h}\footnote{As of today, the numerology of Release 15 supports the speeds of up to 100 km/h \cite{rohdeshwarz2017demystifying}, while higher values correspond to the eMBB use-case requirements and will be addressed in Release 16.}); (ii) massive machine-type communications (mMTC) at high densities (up to one million connections per square km) calling for long battery life, broad range, and ultra-low cost; 
and (iii) ultra-reliable and low latency (URLLC) communications characterized by extremely reliable and available connectivity, high speeds, as well as 1 ms air and 5 ms end-to-end latencies \cite{keysight2017understanding}. %[Keysight tech slides]
%10-20 Gbps, support for high mobility (up to 500mk/h0, energy savings by 100 times
%

Generally, although the 5G NR is defined with band-agnostic operations, which allows this technology to be deployed on any bands without restrictions, 3GPP specifies two major frequency ranges (FR) for Release 15 \cite{mediatek20185GNR}:
\begin{itemize}
\item 450 MHz -- 6 GHz (FR1, referred to as Sub-6 GHz) incorporating bands numbered from 1 to 255,
\item 24.25 GHz -- 52.6 GHz (FR2, commonly referred to as mmWave\footnote{Strictly speaking, mmWave starts at 30 GHz, but the community loosely assigns the slightly lower frequencies to mmWave as well.}) with the bands numbered from 257 to 511.
\end{itemize}
Albeit it is important \textit{not to} misinterpret 5G as a strictly mmWave solution 
%[Olga] deleted comma
since the new standard provides high flexibility and supports a broad range of choices, the mmWave frequencies represent one of the most perspective capabilities of the 5G NR.

Naturally, mmWave communications exhibit certain undeniable advantages including the much wider -- available and yet unoccupied -- bandwidth as well as better spatial reuse and privacy aspects. The latter 
%[Olga]:added 'two'
two are due to the utilization of highly directional transmissions that can be achieved with smaller wavelengths and hence, a higher number of antenna elements. At the same time, the defects of these qualities manifest in higher signal attenuation (including specific atmospheric effects) and implications of clustered multi-path signal structure, which may dramatically increase the bit error rate. Luckily, these negative effects can be mitigated by employing sophisticated beamforming and beam tracking mechanisms that become an indispensable part of NR research and implementation.% such as beamforming and Massive-Multiple Input Multiple Output (M-MIMO)

%"Modern beamforming antenna architectures can help to mitigate these problems by adapting to the channel. This way, delayed multipath components can be ignored or significantly reduced through beam steering.Attenuation Fading Delay Spread"

In this paper, we provide a condensed vision of the 5G NR key features supported by Release 15, which should be taken into account by the engineers and theoreticians while searching for the fundamental trade-offs and evaluating the performance of mmWave-based NR systems, both analytically and through simulation studies. The remainder of this text is organized as follows. Section II outlines the main distinctive features of the NR technology according to Release 15, which boil down in this work to flexible NR numerology and 3D beamforming. Section III outlines the initial access procedure employed by the 5G NR, including a cell search mechanism and a random access procedure. Finally, we conclude with a discussion on open questions and new features expected in Release 16. 

%----------------------------------------------------------------------------------------------------------------------------------------------------------------------------------------------------------------------------------------------------------------------------------%

\vspace{-10px}
\section{5G New Radio Features}\vspace{-5px}

The legacy LTE networks, which could easily be described as a ``one-fits-all" solution, are unable to satisfy the increasingly stringent and highly diverse 5G requirements in terms of reliability, availability, latency, QoS, scalability, and throughput. To support a variety of vertical industries, the 5G NR -- as a global standard for a new \textit{OFDM-based air interface} -- is specifically designed to support a tremendous variety of 5G services and use-cases, device types, and deployments. The officially completed Release 15 ratifies the 5G NR physical layer with an emphasis on constructing flexible \textit{scalable numerology} and \textit{scalable slot duration}. 

In this section, we provide a brief outline of the key 5G NR features with a particular focus on cellular mmWave operation, which has received much attention in the standardization community recently. The most important technical specifications for the purposes of this review are listed in~Table~\ref{table:specs}, whereas the complete list of documents may be accessed online~\cite{3GPPFeatures}. \textcolor{black}{A comprehensive interpretation of the standard may also be found in~\cite{sharetechnote}}. 
We structure the subsequent discussion as two dedicated subsections, one of which addresses the new numerology, while another one elaborates on the new beamforming features supported by the 5G NR.

%\cite{sharetechnote}

%flexible channel bandwidth concept
\begin{table}\vspace{-10px}
\caption{5G NR specification map.}\vspace{-5px}
\begin{center}
\begin{tabular}{|c|l|l|}
\hline
  TS number   &\multicolumn{1}{|c|}{Title} &\multicolumn{1}{|c|}{Version, date$^*$} \\\hline
38.101-1 &UE Radio Transmission and Reception. Part 1.&V15.2.0, 2018-07 \\\hline
%Part 2: Range 2 Standalone
%38.201 &General Description&V15.2.0, 2018-06 \\\hline
38.201 &Physical Layer General Description&V15.0.0, 2018-01 \\\hline
38.202 &Services Provided by the Physical Layer&V15.2.0, 2018-06\\\hline
38.211 &Physical Channels and Modulation&V15.2.0, 2018-06\\\hline
38.212 &Multiplexing and Channel Coding&V15.2.0, 2018-06\\\hline
38.213 &Physical Layer Procedures for Control&V15.2.0, 2018-06\\\hline
38.214 &Physical Layer Procedures for Data&V15.2.0, 2018-06\\ \hline
38.215 &Physical Layer Measurements&V15.2.0, 2018-06 \\\hline
38.300 &Overall Description&V15.2.0, 2018-06\\ \hline
38.321 &MAC Protocol Specification&V15.2.0, 2018-06\\
\hline
\multicolumn{3}{l}{\textsuperscript{*}\footnotesize{Recent version as of July 2018.}}
\end{tabular}
\label{table:specs}
\end{center}
\end{table}

\subsection{New Scalable Numerology and Frame Structure}
\vspace{-1px}

A \textit{numerology} is defined as a set of parameters that specify the OFDM system design and includes Subcarrier Spacing (SCS), Cyclic Prefix (CP), symbol length, and Transmission Time Interval (TTI)\footnote{Also referred to as one slot, multiple slots, or one mini-slot (see below)~\cite{3GPPTS38300}.}. The 5G NR numerology targets various deployments and performance requirements; therefore, it is designed to be configured flexibly to serve diverse purposes. 
%The downlink transmission waveform is conventional OFDM using a cyclic prefix. The uplink transmission waveform is conventional OFDM using a cyclic prefix with a transform precoding function performing DFT spreading that can be disabled or enabled. release15

In particular, one significant difference between the LTE and 5G NR is that the latter defines several SCSs \cite{3GPPTS38213} as opposed to the only option of 15 kHz, which the current LTE standard specifies. Taking 15 kHz as a baseline, the NR numerology is based on the exponentially scalable SCS as defined by $f \textrm{[kHz]} = 15  \cdot 2^\mu$ \cite{3GPPTS38300}, where $\mu$ is referred to as the SCS configuration and takes the values of 0 (15 kHz), 1 (30 kHz), 2 (60 kHz), 3 (120 kHz), or 4 (240 kHz)\footnote{$\mu=-2$ (3.75 kHz), which corresponds to the LTE NB-IoT SCS, is also supported.}. 

In Table~\ref{table:numer}, we collect the range of SCSs that are advised by the current Release 15 as well as provide the respective slot durations and other parameters important for system-level evaluation. We intentionally highlight the \textit{mmWave option}, since this direction remains the primary objective of our paper. Due to the impact of phase noise at higher frequencies, the carrier separation should be increased, which naturally divides our table: the left vertical part belongs to FR1 with narrower bands, while the right part corresponds to FR2, i.e., mmWave frequencies, as also indicated in Table~\ref{table:numer}.

\begin{table}
\vspace{-10px}
\caption{Supported transmission numerologies in 5G NR.}
\begin{center}
\vspace{-20px}
\begin{tabular}{|l|c|c|p{3cm}|c|c|}
\hline
\multicolumn{1}{|c|}{$\mu$} & \multicolumn{1}{|c|}{0} & \multicolumn{1}{|c|}{1}& \multicolumn{1}{|c|}{2}& \multicolumn{1}{|c|}{3}& \multicolumn{1}{|c|}{4}\\
\hline
$\Delta f = 2^\mu \cdot 15 \textrm{ [kHz]}$ & \multicolumn{1}{|c|}{15} & \multicolumn{1}{|c|}{30} & \multicolumn{1}{|c|}{60} & \multicolumn{1}{|c|}{120} & \multicolumn{1}{|c|}{240$^{*}$}
\\
Cyclic prefix$^{**}$ & \multicolumn{1}{|c|}{Normal} & \multicolumn{1}{|c|}{Normal} & \multicolumn{1}{|c|}{Normal, Extended} & \multicolumn{1}{|c|}{Normal} &\multicolumn{1}{|c|}{Normal}
\\
%Cyclic prefix duration& \multicolumn{1}{|c|}{Normal} & \multicolumn{1}{|c|}{Normal} & \multicolumn{1}{|c|}{Normal, Extended} & \multicolumn{1}{|c|}{Normal} &\multicolumn{1}{|c|}{Normal}
%\\
\hline
For data$^{***}$ & \multicolumn{1}{|c|}{+} & \multicolumn{1}{|c|}{+} & \multicolumn{1}{|c|}{+} & \multicolumn{1}{|c|}{+} &\multicolumn{1}{|c|}{--}
\\
For synchronization$^{***}$ & \multicolumn{1}{|c|}{+} & \multicolumn{1}{|c|}{+} & \multicolumn{1}{|c|}{--} & \multicolumn{1}{|c|}{+} &\multicolumn{1}{|c|}{+}
\\
\hline
For data $>$6 GHz& -- & -- & \multicolumn{1}{|c|}{+} & \multicolumn{1}{|c|}{+} &\multicolumn{1}{|c|}{--}
\\
For synch $>$6 GHz& -- & -- & \multicolumn{1}{|c|}{--} & + &\multicolumn{1}{|c|}{+}\\
\hline
Symbol duration, $1/\Delta f$ [$\mu$s]& 66.67& 33.33& \multicolumn{1}{|c|}{16.67} & \multicolumn{1}{|c|}{8.33} &\multicolumn{1}{|c|}{4.17}\\
Slot duration [$\mu$s]& 1000& 500 & \multicolumn{1}{|c|}{250} & \multicolumn{1}{|c|}{125} &\multicolumn{1}{|c|}{62.5}\\
\hline
Number of slots per subframe & 1& 2 & \multicolumn{1}{|c|}{4} & \multicolumn{1}{|c|}{8} &\multicolumn{1}{|c|}{16}\\
Number of slots per frame & 10 &20 & \multicolumn{1}{|c|}{40} & \multicolumn{1}{|c|}{80} &\multicolumn{1}{|c|}{160}\\
\hline
Minimum bandwidth (MHz) \cite{mediatek20185Gdesign}& 4.32 &8.64 & \multicolumn{1}{|c|}{17.28} & \multicolumn{1}{|c|}{34.56} &\multicolumn{1}{|c|}{69.12}\\
Maximum bandwidth (MHz) \cite{mediatek20185Gdesign}& 49.5 &99 & \multicolumn{1}{|c|}{198} & \multicolumn{1}{|c|}{396} &\multicolumn{1}{|c|}{397.44}\\
\hline
Min. number of RBs, UL/DL \cite{sharetechnote}& 24 &24 & \multicolumn{1}{|c|}{24} & \multicolumn{1}{|c|}{24} &\multicolumn{1}{|c|}{24}\\
Max. number of RBs, UL/DL  \cite{sharetechnote}& 275 &275 & \multicolumn{1}{|c|}{275} & \multicolumn{1}{|c|}{275} &\multicolumn{1}{|c|}{138}\\

%Title (centered) &  {\Large\bfseries Lecture Notes} & 14 point, bold\\
%1st-level heading &  {\large\bfseries 1 Introduction} & 12 point, bold\\
%2nd-level heading & {\bfseries 2.1 Printing Area} & 10 point, bold\\
%3rd-level heading & {\bfseries Run-in Heading in Bold.} Text follows & 10 point, bold\\
%4th-level heading & {\itshape Lowest Level Heading.} Text follows & 10 point, italic\\
\hline
\multicolumn{6}{l}{\textsuperscript{$^*$}\footnotesize{480 kHz is not adopted for Release 15~\cite{TS38211}}}\\
 \multicolumn{6}{l}{\textsuperscript{$^{**}$}\footnotesize{Downlink: conventional OFDM with cyclic prefix (CP)}}\\
 \multicolumn{6}{l}{\footnotesize{Uplink: conventional OFDM with CP with optional transform precoding}}\\
 \multicolumn{6}{l}{\footnotesize{CP length is calculated based on slot and symbol length and number of symbols}}\\
  \multicolumn{6}{l}{\footnotesize{per slot (14 for normal, 12 for extended CP~\cite{3GPPTS38300}) }}\\
  \multicolumn{6}{l}{\textsuperscript{$^{***}$}\footnotesize{For either of two frequency ranges}}
\end{tabular}
\label{table:numer}
\end{center}
\vspace{-20px}
\end{table}

% 12 consecutive sub-carriers form a physical resource block (PRB). Up to 275 PRBs are supported on a carrier.

%CHECK LATER
%\textcolor{magenta}{[Giulia]NR frame structure supports Time Division Duplexing (TDD) and Frequency Division Duplexing (FDD) transmissions and operation and introduces three principles to support forward backward compatibility and reduce interactions between different features [28]:
%\begin{itemize}
%\item Self-contained transmissions: data in slot/beam is decodable without dependency on other slots/beams;
%\item Transmissions confined in time and frequency;
%\item Avoid static or strict timing relations across slots and across different transmission directions.
%\end{itemize}
%}

Further, different SCS values are translated into a flexible frame structure. According to Release 15, downlink (DL) and uplink (UL) time is divided into \textit{frames} of 10 ms duration, and each frame comprises ten \textit{subframes} of 1 ms length (both values are constant). The basic transmission unit is a \textit{slot} (TTI), which carries 14 OFDM symbols (or 12 with Extended CP) for SCS of up to 60 kHz and 14 symbols for higher SCSs~\cite{3GPPTS38300}. 
%[Olga] removed 7, added ref
In contrast to LTE, the slot duration can be flexibly modified from 1 ms to 0.0625 ms depending on the selected SCS option (i.e., the duration is calculated as a ratio $1/2^\mu$ ms, see Table~\ref{table:numer}). While shorter slot durations (larger SCSs) aim at supporting low latency and high reliability, longer values (lower SCSs) help increase spectral efficiency and may be suitable for larger cell sizes and thus for the lower frequency ranges as mentioned above.
 
 \begin{figure}[!h]\vspace{-10px}
	\center{\includegraphics[width=1\linewidth]{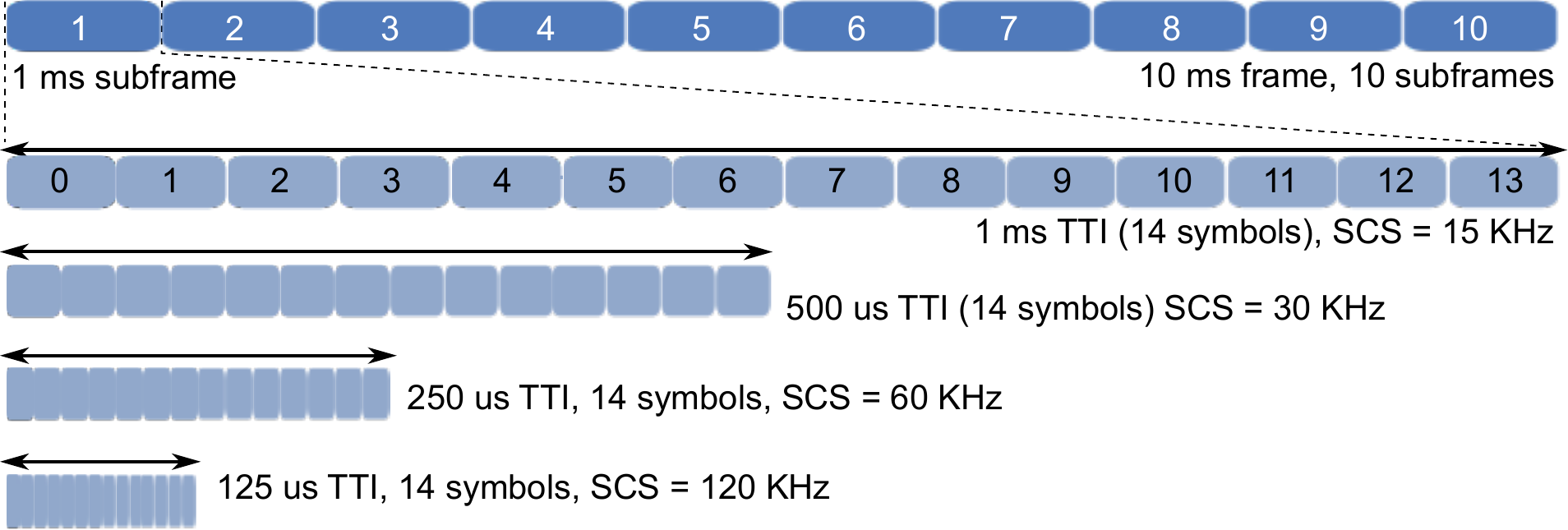}}
	\caption{Scalable NR TTI.}
	\label{fig:1}\vspace{-5px}
\end{figure}

\textcolor{black}{A slot may be used for only DL, only UL, or \textit{mixed} UL and DL transmission (e.g., incorporate both control and data exchange within one slot interval, which may be managed dynamically via a slot format indication -- valid for one or several consecutive slots). This flexibility makes it possible to exchange \textit{TDD self-contained slots}~\cite{TS38802,keysight2017understanding} that incorporate UL/DL scheduling, data, and acknowledgment all at once and represent one of the key enablers for URLLC. }
\textcolor{black}{Another new NR entity beneficial for URLLC is \textit{a mini-slot} (minimal schedulable resource optimized for short data transmissions), which may occupy 2, 4, or 7 OFDM symbols and start at any time without waiting for the slot boundary~\cite{Qualcomm2018Designing}. Release 15 also supports \textit{slot aggregation}, so that the data may be scheduled over multiple slots~\cite{mediatek20185GNR}, even over slots having different formats~\cite{TS38211}. }
%can become a key solution for the low latency scenarios
%Both FDD and TDD are supported.!!!

%Another revolutionary feature of NR is introducing a \textit{mini-slot}~\cite{Zaidi2017}, which is defined as a minimum schedulable resource and can be as short as one OFDM symbol (for at least FR2) starting anytime. 
%\todo[]{I have concerns: are mini-slot and self-contained subframe the same?}
%
%
%[1] 3GPP TSG RAN WG1 Meeting NR#3 : R1- 1716650 Comparison of PBCH DMRS mapping schemes
%[2] 3GPP TSG RAN WG1 Meeting NR#3 : R1-1715841 Remaining Details on PBCH design and contents
%[3] 3GPP TSG RAN WG1 Meeting AH_NR#3 : R1-1716609 - On remaining details of NR DL DMRS
%[4] 3GPP TSG RAN WG1 NR Ad-Hoc#3 : R1-1716574 - Discussion on time domain resource allocation

\subsection{Directivity and Beamforming at FR2}

%:Modern beamforming antenna architectures can help to mitigate these problems by adapting to the channel. This way, delayed multipath components can be ignored or significantly reduced through beam steering. Antennas that are designed to adapt and change their radiation pattern in order to adjust to the RF environment are called active phased array antennas"
Another distinctive feature of 5G NR at mmWave frequencies is the possibility to rely on beam steering by highly directional antennas, which has become feasible for a wide range of use-cases due to smaller antenna elements and larger antenna arrays. Moreover, the use of highly directional antennas at the NR base station (gNB) and/or at the user equipment (UE) represents a natural solution to compensate for faster signal attenuation and improve the link budget. Importantly, beamforming is not an exclusive mmWave-specific feature -- it can also be used at lower frequencies; however, when it comes to extremely high frequency range, beamforming becomes the only viable choice for most of the envisioned use-cases.

%"structure of NR would be merely a hollow place holder. MIMO techniques help to differentiate NR from 4G in many aspects. NR would operate up to 100 GHz bands. At such high carrier frequency, MIMO or beamforming is a must⁃have fea⁃ ture. Otherwise, the severe path loss and penetration would render NR useless, even indoor. Hence, beamformed transmis⁃ sion would widely be employed, not only for traffic channels, but also for control signaling, random access signal, synchroni⁃ zation signal, and broadcast channels carrying system informa⁃ tion. Beamforming is also used at the receiver. Because of these, initial access and frame structure of NR would have many footprint of beamforming, which is seldom seen in LTE." [5G New RAdio]

In general, beamforming techniques are responsible for controlling the properties of electromagnetic radiation patterns and the gain of an antenna array by aligning the amplitude and phase of transmit/receive signals. By doing so, a device is able to form an appropriate beam-pattern by increasing the antenna gains toward the desired direction and, at the same time, suppressing the radiation sideways and changing the interference footprint. 

The most usable beamforming algorithms include exhaustive search (brute-force sequential beam searching over a predefined codebook, which is a set of beams to multiple directions covering the entire angular space) and iterative or hierarchical search (two-stage scanning, which transmits the signals over wide sectors and then refines within the best sector by steering narrower beams)~\cite{giordani2016initial,desai2014initial}. However, in case of narrow beams, these simple solutions may likely result in excessive delays and can be inefficient for certain use-cases. To properly align the beams within a limited delay budget, devices might need to exploit alternative intricate techniques for beam and mobility management that currently generate a particular interest within the research community.

Generally, beamforming may be categorized as two- or three-dimensional:%\vspace{-2px}

\begin{itemize}
\item 2D beamforming: controls the radiation pattern in one plane (the antenna elements are positioned as a linear array).%employs linear array antennas,
\item 3D beamforming\footnote{also known as \textit{elevation beamforming}}: steers the antenna beams not only in the azimuth but also in the elevation plane (planar flat/volume arrays).%implemented based on planar flat/volume array antennas
\end{itemize}

Compared to the conventional 2D techniques, 3D beamforming in 5G NR is built on up to 256 (32) antenna elements for gNB (UE) and supports vertical sectorization with an additional sector division in the radial direction. This extra sectorization allows reducing co-channel interference and creating a higher degree of freedom in optimizing the system performance without altering the existing physical architecture~\cite{razavizadeh2014three}. We note that 3D beamforming is supported not only by the NR but also by other mmWave technologies, such as IEEE 802.11ad/ay~\cite{ieee80211ay}. The beamforming architectures included \textcolor{black}{in} the 5G NR are~\cite{docomo2018}:

\begin{itemize}
\item \textit{Analog beamforming}: exploits a single RF chain and multiple phase shifters, which results in simpler beam-search procedures (one beam at a time). Characterized by low power consumption and low complexity (used in, e.g., IEEE 802.11ad~\cite{80211ad}).
\item \textit{Digital beamforming}: requires several RF chains, one for each antenna, and thus is able to support multi-stream operation (e.g., MU-MIMO). Characterized by higher flexibility in shaping the beams but also by increased costs, power consumption, and complexity (used in, e.g., LTE and supported by IEEE 802.11ay~\cite{ieee80211ay}).
\item \textit{Hybrid beamforming}: represents a compromise between the analog and digital options based on dividing the precoding between the analog and digital domains. Characterized by fewer supported streams than digital, lower complexity, and lower power consumption due to a decreased number of RF chains. 
\end{itemize}

Hybrid beamforming constitutes a relatively recent solution to combine the strengths of both options above and promises nearly the same performance as achieved by the digital beamforming~\cite{alkhateeb2014mimo}.
%low power consumption and low complexity. For narrow-beam Analog Beamforming requires several RF chains to serve UEs that are separated geographically. 

%\textcolor{red}{[picture with hybrid BF?]}

%Hybrid Beamforming is the trade-off between Analog and Digital Beamforming: digital precoding is used for the effective channel consisting of the analog beamforming weights and the actual channel matrix. 

%%, a set of Layer 1/Layer 2 procedures to acquire and maintain a set of gNB or UEs beams, plays 
%The key role belongs in NR belongs to the beam management and includes the following aspects:
%\begin{itemize}
%\item beam \textit{determination} to select the best Tx/Rx beams,% (at gNBs or UEs),
%\item beam \textit{measurement} to measure the characteristics of the signals detected in the previous procedure,% (at gNB and UEs), 
%\item beam \textit{reporting} to report information relative to measured signals,% (at UEs),
%\item beam \textit{sweeping} to cover the entire spatial area with beams transmitted/received sequentially.% in a pre-determined way.
%\end{itemize}

%The . For these bands, the plan is to extend Release 13 and Rel14 FD-MIMO framework to support 64, 128, or even 256total physical elements, with flexible CSI acquisition and beamforming.

%to benefit from multiplexing to increase data-rate or user spatial diversity to achieve robustness to blockage

Finally, with respect to directional data transmission, the NR standard enables different design options for MIMO systems (up to 256 antenna elements). To increase the data rate and improve the spatial diversity, 5G NR supports eight streams for the single-user MIMO operation and twelve streams for the multi-user MIMO in DL, as well as four streams for the single-user MIMO operation in UL~\cite{docomo2018}. Since MIMO functionality requires continuous evaluation of the channel quality, it may also benefit from utilizing the self-contained subframe structure described above by transmitting the UL control information and sounding reference signals~\cite{Qualcomm2018making}.
%Figure 16 shows an example of a TDD downlinkcentric subframe, where data transmission is from the network to the device and the acknowledgement is sent by the device back to the network in the same subframe.@  making-5g-a-real

%[1]3GPP R1-166089. 3GPP TSG RAN WG1 Meeting #86 - Beam Management Procedure for NR MIMO
%[2] 3GPP R1-166214. 3GPP TSG RAN WG1 Meeting #86 - Discussion on the beam management for the NR
%[3] 3GPP R1-166389. 3GPP TSG RAN WG1 Meeting #86 - Beam Management in Millimeter Wave Systems
%[4] 3GPP R1-166565. 3GPP TSG RAN WG1 Meeting #86 - Beam management without prior beam information
%[5] 3GPP R1-166657. 3GPP TSG RAN WG1 Meeting #86 - Views on beam management for NR
%[6]3GPP R1-166785. 3GPP TSG RAN WG1 Meeting #86 - Discussion on TRP beamforming and beam management
%[7]3GPP R1-167466. 3GPP TSG RAN WG1 Meeting #86 - Key principles for beam management
%[8] 3GPP R1-167467. 3GPP TSG RAN WG1 Meeting #86 - Reference signals and reports to support beam management
%[9] 3GPP R1-167543. 3GPP TSG RAN WG1 Meeting #86 - Beam Management Considerations for above 6 GHz NR

%----------------------------------------------------------------------------------------------------------------------------------------------------------------------------------------------------------------------------------------------------------------------------------%

\section{Initial Access in 5G NR}

\begin{figure}[!h]\vspace{-10px}
	\center{\includegraphics[width=0.9\linewidth]{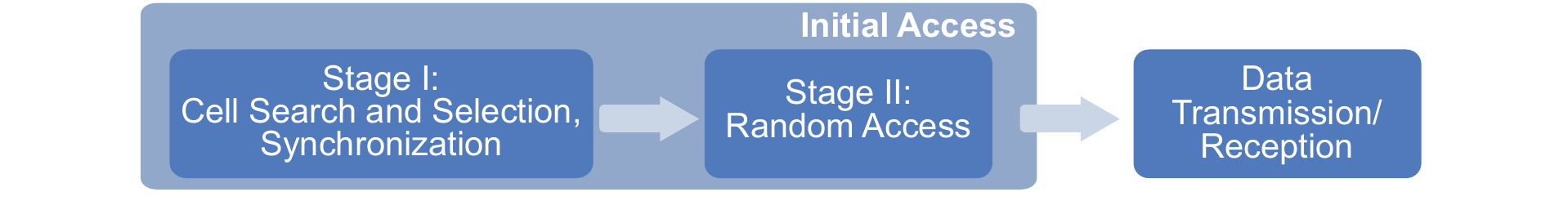}}\vspace{-5px}
	\caption{Steps of the initial access procedure.}
	\label{fig:2}\vspace{-15px}
\end{figure}

Generally, initial access in cellular systems comprises several consecutive steps, which we broadly divide into two stages (see Fig.~\ref{fig:2}): 
\begin{itemize} 
\item Stage I: cell search and synchronization (acquiring system information). 
\item Stage II: random access procedure.
\end{itemize} 

In 5G NR, initial access resembles a standard procedure that the legacy LTE relies upon: in particular, it includes receiving synchronization signals, extracting system information, and establishing a connection via a random access procedure. However, regarding how the initial access is performed for FR2, 5G NR differs from LTE operation significantly, which is primarily due to its highly directional nature at the frequencies above 6 GHz. 
\vspace{-5px}

A key difference emerges already at Stage I during the cell search and synchronization. We remind that in LTE the synchronization signals are transmitted by using omnidirectional antennas (beamforming, if employed, applies only after synchronization, i.e., during the data transmission). In contrast, to extend communication distances, an NR gNB employs beam sweeping\footnote{A sequential transmission by using the entire codebook or its subsets.} already when broadcasting the synchronization signals. Hence, initial access for mmWave NR becomes much more challenging, since both the UE and the gNB have to detect the correct directions and align the beams before subsequent data transmission. 
%by concentrating transmit signal power in a specified direction, 
%SSBs need to reach all UEs within a cell, so  is supported for multi-beam operations and is applied to transmit SSBs. 
%For frequencies below 6 GHz is possible a single-beam approach. 

%, as well as more sophisticated alternatives . 

\vspace{-5px}
\subsection{Stage I: Cell Search and Selection}

\begin{figure}[!h]
\vspace{-10px}
	\center{\includegraphics[width=.9\linewidth]{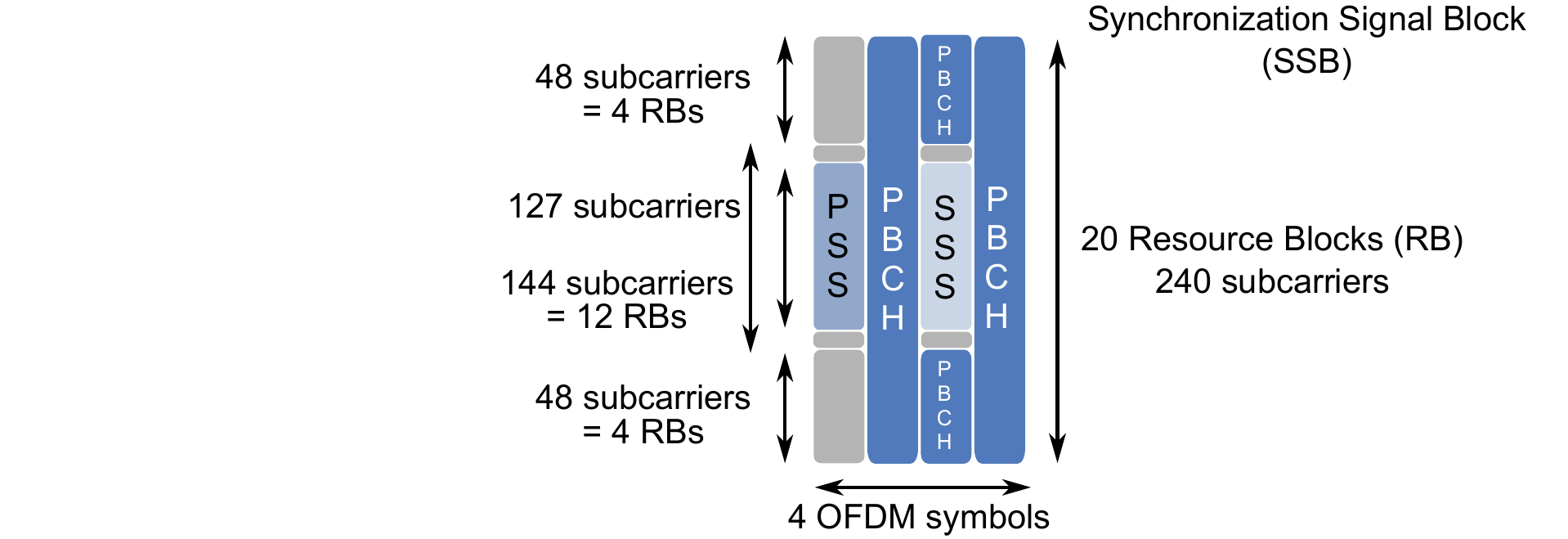}}
	\caption{Time-frequency structure of the synchronization signal and PBCH block (consists of PSS and SSS, which cover 1 OFDM symbol and 127 subcarriers (SCs) each, and PBCH that occupies 3 OFDM symbols and 240 SCs, respectively). One Resource Block (RB) = 12 SCs.}
	\label{fig:3}
	\vspace{-10px}
\end{figure}

At Stage I, the gNB periodically broadcasts \textit{Synchronization Signal Blocks}\footnote{Also referred to as the synchronization signal and PBCH block in the specifications, but may be simply understood as a beacon.} (SSBs), which contain (i) Primary Synchronization Signal (PSS), (ii) Secondary Synchronization Signal (SSS), and (iii) Physical Broadcast CHannel (PBCH)~\cite{3GPPTS38300,TS38211} as demonstrated in Fig.~\ref{fig:3}. In contrast to LTE, the synchronization signals and PBCH (carries system information) are inseparable in 5G NR. Each SSB is mapped onto a \textit{different beam} and broadcasted by the gNB to its proximate UEs. 

A cell search procedure is used by the UE to acquire time and frequency synchronization with the cell as well as to determine the Cell ID. The UE listening on a channel first detects the symbol timing and Physical Cell ID (PCI) in PSS over the time domain. Then, by utilizing SSS, the UEs obtain information regarding the frame timing in the frequency domain, CP length, as well as detect FDD/TDD and acquire the reference signals for demodulation~\cite{mediatek20185Gdesign}. 

Importantly, the periodicity of the SSB is configured by the network, while the default transmission periodicity, which is assumed by the UE before such notifications, is 20 ms (i.e., 2 NR frames). This interval is four times longer than that in LTE (5 ms) and aims at reducing the ``always-on" transmission overheads. The frame and slot timings are defined by the identifiers of SBSs and acquired by the UE as described above. 
%the time locations where the gNB sends SSBs are automatically defined by the current SCS. 

More specifically, SSBs may be transmitted in a batch by forming an \textit{SS Burst} (one SSB per beam) that may be used during beam sweeping; a collection of SS Bursts is referred to as an \textit{SS Burst Set}. Both SS Burst and SS Burst Set may contain one or more elements, while the maximum number of SSBs in an SS Burst is frequency-dependent and takes the values of 4 (below 3 GHz), 8 (3 to 6 GHz), or 64 (6 to 52.6 GHz)~\cite{rohdeshwarzslides}. 

\begin{figure}[!h]\vspace{-15px}
	\center{\includegraphics[width=1\linewidth]{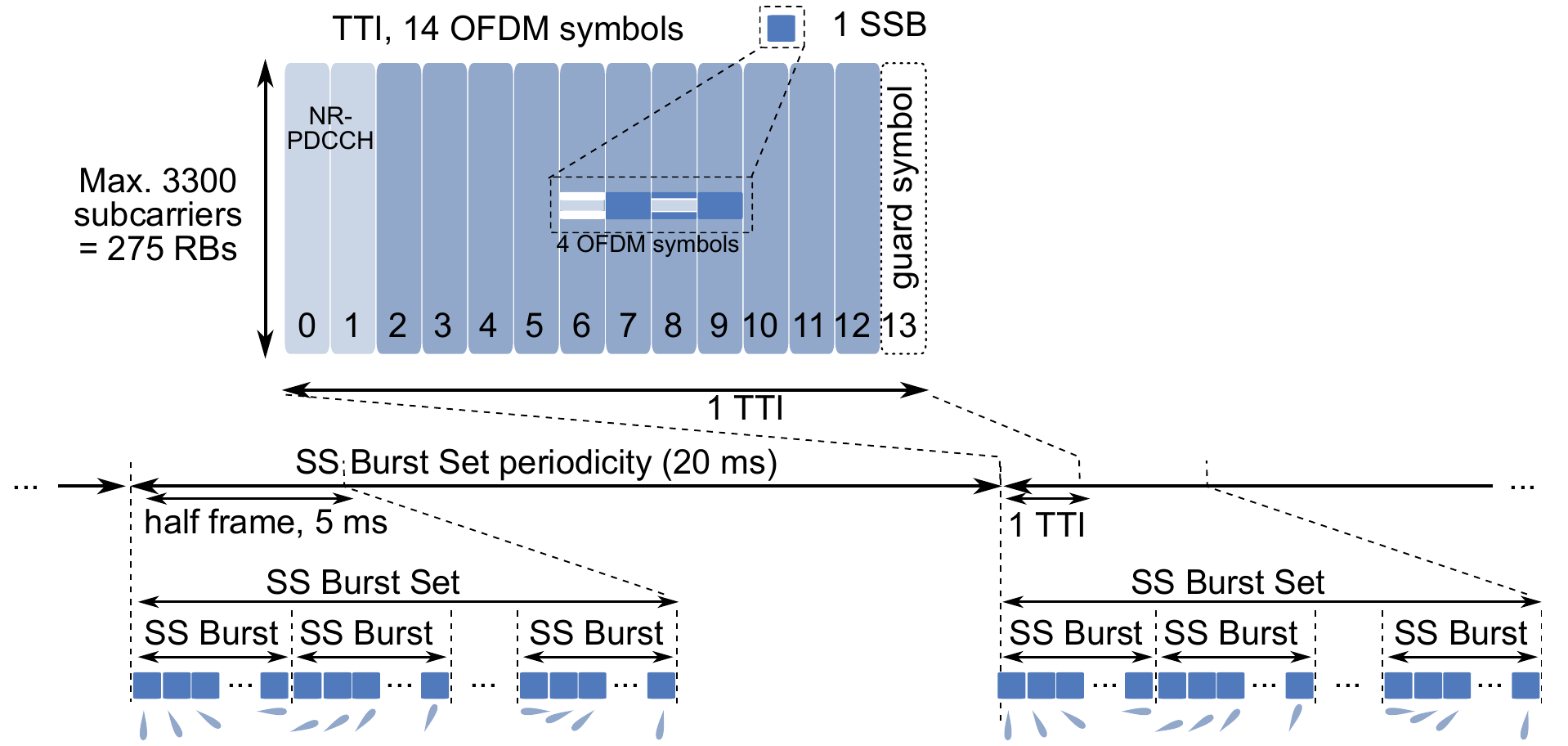}}
	\caption{SSB and SS burst composition~\cite{rohdeshwarzslides}. One SSB corresponds to one beam. }
	\label{fig:4}\vspace{-15px}
\end{figure}

In Fig.~\ref{fig:4}, we illustrate the concept of SS bursts as well as outline the structure of one TTI and demonstrate the share of resources occupied by one SSB. \textcolor{black}{SS Burst Set may occupy one half frame (5 ms), 
%[olga] "half frame" is official
and the beam pattern repeats every 2 frames~\cite{TS38213} (by default for the initial access)}. The overheads created by the occupied resources may be calculated based on the number of SCs (up to 3300), the size of the SSB (addressed in Fig.~\ref{fig:3}), and the number of SSBs per a time unit, which is defined by the required number of beams and their periodicity.

%Importantly, in NR the periodicity and timing of synchronization signals transmission are configured by network. 

The gNB may define multiple candidate positions for SSBs within a radio frame, and this number corresponds to the number of beams. Identification of which SSB is detected and thus acquisition of the frame and slot timings is facilitated by DeModulation Reference Signal (PBCH DMRS), which plays the role of a reference signal for decoding PBCH instead of Cell Specific Reference Signal (CRS) used in LTE.

%. To overcome this issue, one can take advantage of PBCH and Reference Signal (PBCH DeModulation Reference Signal) included in the SSB \cite{docomo2018}. 
% (PBCH-DMRS)

%FIXME \textcolor{red}{[LTE-NR coexistence?]}
%In NR, the number of Physical Cell IDentifiers (PCIDs) expressed by the synchronous signals is doubled w.r.t. LTE since a deployment scenario potentially with extremely high density is supposed. 

%After the successful execution of the cell search procedure above described, the UE can receive the Physical Broadcast Channel (PBCH) can read the Master Information Block (MIB) and System Information Blocks (SIB). The MIB carries the most important system information, such as downlink bandwidth and system frame number, while SIB carries information about Public Land Mobile Network (PLMN), Cell Identity and Cell Status. 

%Cell Search and Selection is an essential procedure and the basic of every interaction between UEs and Network. Cell Search is also used during LTE mobility processes. The UE measures signals threshold based on Reference Signal Received Power (RSRP) (as well as Reference Signal Reported Quality (RSRQ)) and reports them to the serving eNB. RSRP represents the average of the power of all resource elements which carry cell-specific reference signal over entire bandwidth, while RSRQ represents the ratio between the RSRP and the Received Signal Strength Indicator (RSSI). 

\subsection{Stage II: Random Access}

%\cite{lien20175g},%5G new radio: Waveform, frame structure, multiple access, and initial access.
%\cite{R11609117},%Discussion on RA pro-cedure

A random access procedure in 5G NR may be triggered by, e.g., handover, initial access from idle/inactive modes, or beam failure recovery, and usually falls into either of the two categories: contention-free (CFRA) and contention-based random access (CBRA). Here, we focus on the latter option. % as the former may not be of interest from the directivity perspective.
As mentioned above, random access in mmWave cellular generally shares most of its functionality with LTE, which is built upon Physical Random Access CHannel (PRACH) preamble considerations; however, the nature of high directivity imposes new challenges and allows further options. Since single-beam operation (corresponding to omnidirectional transmission/reception at the frequencies below 6 GHz) is similar to LTE RACH functionality, we concentrate on multi-beam operation that arises from using directional antennas~\cite{R11609117}.
%
%When we refer to the directional transmission, the main challenge is to establish the best beam alignment between the transmitter and receiver. First, synchronization signals (SSBs) are transmitted through multiple Tx beams sweeping over the entire angular space. 

At the beginning of the random access procedure, both the UE and the gNB are not aware of the appropriate beam directions; hence, the initial access signals are sent via multiple Tx beam sweeping. After detecting the initial synchronization signals, the UE is able to select the best gNB Tx beam for further DL data acquisition from the beams used to transmit the initial access signals. The gNB also utilizes multiple Rx beams to cover the entire angular space, since the position of this potentially attempting UE is unknown. The gNB provides multiple RACH resources (SSBs) to the UE and applies one Rx beam per each RACH resource that it announced previously. The number of RACH resources within one RACH occasion\footnote{Time allocated for sending preambles.} and the number of contention-based preambles per one beam are also advised by the serving gNB (the maximum is 64 preambles~\cite{TS38211}).

%%One way to reduce the required beam training time during RA procedure is to use multiple beams simultaneously. It is obvious that transmitting/receiving RACH preamble in multiple directions at the same time is beneficial at the cost of more gNB/UE power consumption. !!!

%\cite{R11609118},%,RACH design with and without beam reciprocity
%STI

Importantly, the design of the random access procedure may vary depending on the presence of Rx/Tx reciprocity at the UE or the gNB (no reciprocity, partial, or full reciprocity)~\cite{R11608966,R11609118}. If the gNB relies on beam reciprocity, it maps the UL RACH resources onto the DL initial access signals before the UE starts the RACH procedure~\cite{TS38213} and by that may significantly reduce the required beam training time. 

In general, the NR RACH procedure includes the following four steps~\cite{R11609117} (tailored here to the case of full reciprocity for simplicity~\cite{Yifei2017}): 
\begin{enumerate}
\item Based on the synchronization information from the gNB, the UE selects a RACH preamble sequence (MSG1) and sends it at the nearest RACH occasion (occurs every 10, 20, 40, 80, or 160 ms). Due to reciprocity, the UE may use the Tx beam corresponding to the best Rx beam determined during synchronization. If reciprocity is available at the gNB, the UE transmits only once; otherwise, it repeats the same preamble for all the gNB Tx beams.
\item The gNB responds to the detected preambles with a random access response (RAR) UL grant (MSG2) in PDSCH by using \textit{one selected} beam. After that, the UE and the gNB establish coarse beam alignment that could be utilized at the subsequent steps.
%After this, the TRP and the UE could build a coarse beam alignment, which can be used for the subsequent RACH message transmission and also for the system information delivery after the RACH procedure.
\item Upon receiving MSG2, the UE responds by MSG3 over the resources scheduled by the gNB, which is thus aware where to detect the MSG3 and which gNB Rx beam should be used.
\item The gNB confirms the above by sending MSG4 in PDSCH using the gNB Tx beam determined at the previous step. 
\end{enumerate} 

Without the beam reciprocity, the UE transmits identical MSG1 signals with the same UE TX beam during one RACH occasion, while the gNB receives the preamble by the gNB Rx beam sweeping. The UE changes its Tx beam at the next RACH occasion. Due to repeated transmission, preambles do not need a CP and a guard period; they hence shorten compared to the reciprocity case. Moreover, MSG2 also needs to be sent by sweeping the gNB Tx beams (unless the UE informs the gNB regarding the best gNB Tx beam). The time required for achieving beam alignment varies depending on the codebook length (i.e., the number of combinations of gNB beams and UE beams). We note that as full sweeping might significantly increase the delay, another potential solution for the UE is to act as if reciprocity holds; then in case of a failure the gNB may request to retransmit multiple preambles.
%A trade-off exists between these two method and study is needed to identify the issues.

%
%TRP can receive RACH symbol from each UE with TRP Rx beam index sweeps and detect RACH symbol when TRP Rx beam index and UE Tx beam index is aligned. It is marked with grey color in figure 4. Even if multiple UEs are trying to transmit RACH preamble, orthogonality between multiple UEs with different Rx beam index can be maintained thanks to consecutive RACH symbols. However the number of available multiplexing UEs may be decreased comparing to the scenario with beam reciprocity. In order to support flexible RACH coverage, TRP can use one RACH symbol or combine multiple RACH symbols. TRP should inform UE that how many repetitions would be needed for RACH transmission corresponding to the number of TRP Rx beam indices via SIB[4]. Thus, single preamble format for RACH symbol can be considered represented in Table 2. 

\begin{figure}[!h]
\vspace{-20px}
	\center{\includegraphics[width=0.9\linewidth]{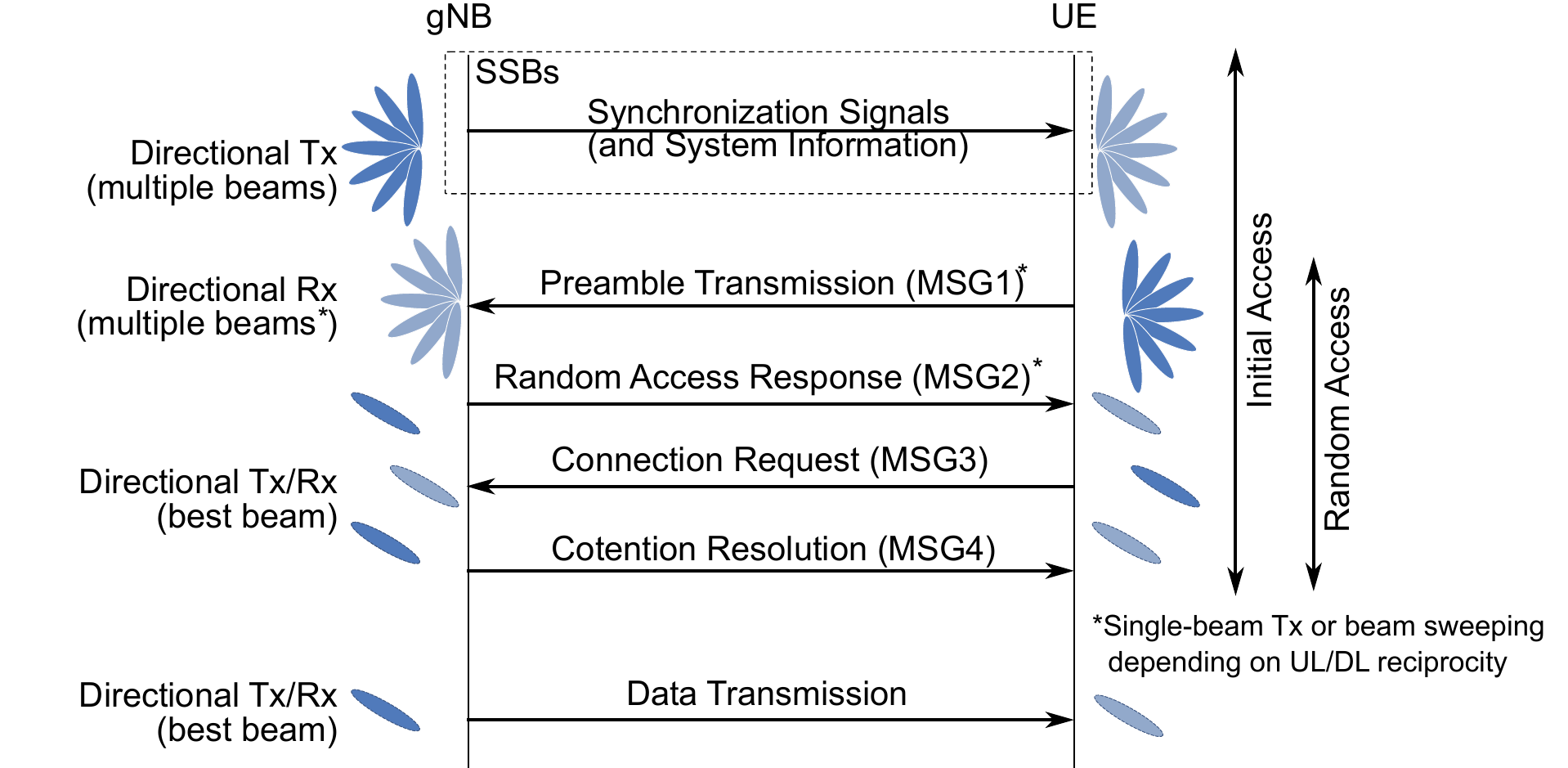}}
	\caption{An example of 5G NR CBRA procedure: no preamble collisions, reciprocity is not available at the UE (MSG1 is sent by multiple beams). }
	\label{fig:random_access}\vspace{-20px}
\end{figure}

%The super wide system bandwidth means that multiple numerologies may coexist in FDM fashion. Ideally for numerology, one synchronization signal would be defined. However, this may lead to excessive overhead. Right now, RAN1 is striving for minimizing the number of subcarrier spacings for synchroni⁃"

If two or more UEs select the same preamble, it may be decoded at the gNB as one preamble, and the gNB then transmits its RAR as for one UE. In this case, a preamble collision occurs at the third step above, when the UEs transmit their requests by using the same resources and perceive a preamble failure after the contention resolution timer expires (instead of receiving MSG4). After a collision, the UE triggers a backoff time, which is selected randomly based on the backoff window (ranges from 5 to 1920 ms~\cite{TS38321}), and restarts. The UE transmits with its default power or the power advised by the gNB. In case of an unsuccessful transmission, the UE follows a power ramping procedure~\cite{TS38321} similar to LTE power ramping (while the power does not change during sweeping).   

\begin{figure}[!h]\vspace{-10px}
	\center{\includegraphics[width=0.8\linewidth]{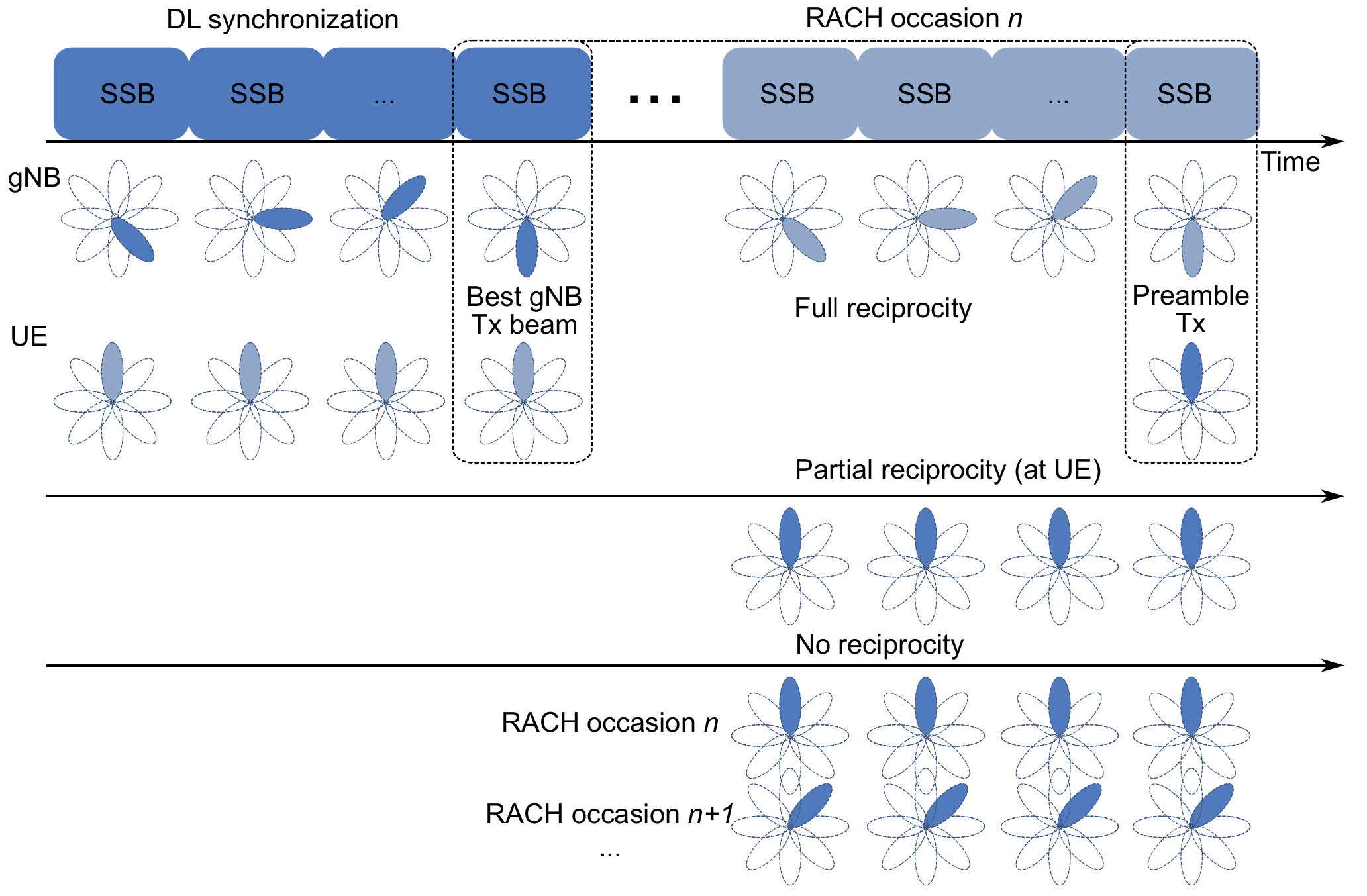}}\vspace{-5px}
	\caption{Example of RACH operation with full/partial/no reciprocity.}\vspace{-15px}
	\label{fig:random_access}
\end{figure}

\begin{table}
\begin{center}%\vspace{-1px}
\begin{threeparttable}
\caption{Antenna configuration options}\vspace{-10px}
\small
\label{table:NRvsLTE}
\begin{tabular}{|*{1}{L{\dimexpr0.3\linewidth-8\tabcolsep\relax}}@{}|*{1}{L{\dimexpr0.38\linewidth-8\tabcolsep\relax}}@{}|*{1}{L{\dimexpr0.38\linewidth-8\tabcolsep\relax}}@{}|}
\hline%\toprule
\multicolumn{1}{|c|}{\bfseries Configuration}&
\multicolumn{1}{|c|}{\bfseries Strengths} &
\multicolumn{1}{|c|}{\bfseries Weaknesses}  \\
\hline%\midrule
Omnidirectional Tx -- Omnidirectional Rx (\textit{Fully-Omnidirectional})
&
Fast, has low complexity, lowest overhead &
High signal attenuation, not suitable for most mmWave applications\\
\hline
Directional Tx -- Directional Rx (\textit{Fully-Directional})
&
Compensated signal attenuation, wide coverage area, high throughput, reduced preamble collision probability
&
Challenging cell search, high complexity, high latency, accentuated deafness and blockage phenomena
\\
\hline
Directional Tx -- Omnidirectional Rx (\textit{Semi-Directional})
&
Reasonable compromise, low complexity, low latency &
Subject to mmWave link instability
 \\
\hline
% \multicolumn{2}{l}{\textsuperscript{*}\footnotesize{Detailed comparison between LTE in terms of numerology and channels may be found in \cite{mediatek20185Gdesign}.}}
\end{tabular}
\end{threeparttable}
\label{table:antenna}
\end{center}\vspace{-20px}
\end{table}

\subsection{Antenna Configurations}

Given the two alternatives -- onmidirectional and directional transmission in 5G NR -- we may differentiate between three possible antenna configurations that can be used during the initial access phase, namely, fully-omnidirectional (OD), fully-directional (FD), and semi-directional (SD) (see Table~\ref{table:antenna}):
\begin{itemize}
\item Omnidirectional Transmissions -- Omnidirectional Receptions (FO),
\item Directional Transmissions -- Directional Receptions (FD),
\item Directional Transmissions -- Omnidirectional Receptions (SD).
\end{itemize}

Here, the first configuration (FO) is the least preferable solution for most mmWave scenarios (except for short line-of-sight links), because a fully omnidirectional link suffers from uncompensated high path loss and thus the coverage area is reduced significantly; however, it allows for much faster and simpler beam search. The second configuration is the most beneficial in terms of the signal-to-noise ratio and throughput but requires complex beam search algorithms and causes delays as the UEs perform an entire angular-space scanning; however, when utilized during the RACH preamble transmission, it could significantly reduce preamble collision probabilities in dense scenarios. 

Finally, the third configuration may become the desired compromise and successfully aid during the synchronization procedure, which requires minimizing delays at most. While directional transmission allows compensating the decreased link budget, omnidirectional reception reduces the complexity and delay of the beam search procedure; however, this option may also lead to the coverage area mismatch. A choice between these three options can be based on the use-case requirements, which include, e.g., link length, UE density, target preamble collision probability, latency, and/or complexity of the device.

%\begin{figure}[!h]
%	\center{\includegraphics[width=\linewidth]{beamforming_conf.png}}
%	\caption{....}
%	\label{fig:1}
%\end{figure}

\begin{table}
\vspace{-15px}
\begin{center}
\begin{threeparttable}
\caption{Selected LTE and NR differences$^*$\vspace{-15px}}\vspace{-5px}
\small
\label{table:NRvsLTE}\vspace{-2px}
\begin{tabular}{|*{1}{L{\dimexpr0.53\linewidth-8\tabcolsep\relax}}@{}|*{1}{L{\dimexpr0.53\linewidth-8\tabcolsep\relax}}@{}|}
\hline%\toprule
{\bfseries LTE features} &
{\bfseries NR features}  \\
\hline%\midrule
Microwave frequencies & 
Two available frequency ranges: below 6 GHz and above 6 GHz (mmWave)\\
\hline%\midrule
Fixed physical layer parameters & 
Scalable physical layer parameters (e.g., flexible SCS, scalable TTI)\\
\hline%\midrule
Synchronization signals are separately and omnidirectionally transmitted & 
Synchronization signals are grouped in SSBs and directionally transmitted\\
%\multicolumn{1}{L{\dimexpr0.5\linewidth-4\tabcolsep\relax}}{} & \\
\hline%\midrule
Cell search procedures are fast and have low complexity &Complex cell search procedures\\
\hline
Random access procedures have higher preamble collision probability &Directional preamble reception may reduce preamble collision probability during random access\\
\hline%\bottomrule
90\% bandwidth efficiency (100 RB cover 18 MHz of 20 MHz bandwidth carrier)&Higher bandwidth efficiency reaching 99\% \\
\hline
Direct Current (DC, with no information) subcarrier helping to locate the frequency&No explicit DC subcarrier reserved for downlink nor uplink\\
\hline
 \multicolumn{2}{l}{\textsuperscript{*}\footnotesize{\textcolor{black}{Detailed LTE/NR comparison w.r.t. numerology and channels may be found in~\cite{mediatek20185Gdesign}.}}}\\
\end{tabular}
\end{threeparttable}\vspace{-5px}
\end{center}
\vspace{-10px}
\end{table}

%o FD requires complex algorithms for cell search and a lot of time, but allows longdistance communications and high throughput, could be a good solution for distant UEs

%o SD does not require complex algorithms and it is a fast procedure, but the coverage areais significantly reduced, could be a good solution for UEs close to the gNB

%• FD mode at gNB could be a good solution to reduce preamble collision probability, while SD could be a good solution to reduce cell search latency and complexity

%\cite{TR38912},%release 14
%\cite{TR138901},%Study on channel model for frequencies from 0.5 to 100 GHz (3GPP TR 38.901 version 14.3.0 Release 14)
%\cite{TS23501},%3GPP TS 23.501: System Architecture for the 5G System (Jul 2017)

%\cite{Qualcomm2018Expanding},%Qualcomm: Expanding the 5G NR ecosystem: 5G NR roadmap in 3GPP Rel 5G
%\cite{Zaidi2017}%New Radio: designing for the future. Ericsson
%\todo[]{put where they belong}

%----------------------------------------------------------------------------------------------------------------------------------------------------------------------------------------------------------------------------------------------------------------------------------%

%[31] I. Takao, “5G Standards Progress and challenges” Radio and Wireless Symposium, pp 1.4, 2017.
\vspace{-10px}
\section{Conclusion}\vspace{-5px}

Our short review of the key differences between 5G NR and LTE is summarized in Table~\ref{table:NRvsLTE}. More specific \textcolor{black}{numbers} 
%figures 
in terms of the physical layer procedures are available in~\cite{mediatek20185Gdesign}. While Release 15 addresses the most essential features to deploy 5G networks, the following Release 16 -- planned to be completed in December 2019 -- targets to maintain the full-fledged 5G vision. The study items listed for Release 16~\cite{3GPPFeatures} as of today include the following developments beyond Release 15: NR-based access to unlicensed spectrum (to be studied for both licensed-assisted access and stand-alone deployments, similarly to LAA), Non Orthogonal Multiple Access (NOMA, allowing multiple UEs to access the channel over the same resources by relying on multi-user detection algorithms), evaluation of advanced V2X use-cases for NR and LTE, backhauling options for NR, industrial IoT scenarios, 5G-grade URLLC enhancements, solutions to support non-terrestrial networks, dual connectivity enhancements, and 5G for satellites.

\vspace{-7px}
\bibliographystyle{splncs04}
\bibliography{giuliabib}

\begin{thebibliography}{10}
\providecommand{\url}[1]{\texttt{#1}}
\providecommand{\urlprefix}{URL }
\providecommand{\doi}[1]{https://doi.org/#1}

\bibitem{sharetechnote}
\url{http://www.sharetechnote.com}, accessed in June 2018

\bibitem{mediatek20185Gdesign}
{5G Design Concepts Towards the Next Generation Networks}. White Paper (2018),
  \url{https://cdn-www.mediatek.com/page/5G_Design_Concepts_v12_MARCOM.pdf}

\bibitem{mediatek20185GNR}
{5G NR -- A New Era for Enhanced Mobile Broadband}. White Paper (2018),
  \url{https://cdn-www.mediatek.com/page/MediaTek-5G-NR-White-Paper-PDF5GNRWP.pdf}

\bibitem{3GPPTS38300}
{3GPP Technical Specifications Group RAN}: {TS 38.300: NR and NG-RAN Overall
  Description, Stage 2 (Release 15)} (Dec 2017)

\bibitem{TS38211}
{3GPP Technical Specifications Group RAN}: {TS 38.211 (Release 15): Physical
  channel and modulation} (June 2018)

\bibitem{3GPPTS38213}
{3GPP Technical Specifications Group RAN}: {TS 38.213: NR Physical layer
  procedures for control (Release 15)} (June 2018)

\bibitem{TS38213}
{3GPP Technical Specifications Group RAN}: {TS 38.213 (Release 15): Physical
  Layer Procedures for Control} (June 2018)

\bibitem{TS38321}
{3GPP Technical Specifications Group RAN}: {TS 38.321 (Release 15): Medium
  Access Control (MAC) protocol specification} (June 2018)

\bibitem{R11608966}
{3GPP Technical Specifications Group RAN}: {R1-1608966: Considerations on
  Sweeping Time Interval in NR} (Oct 2016)

\bibitem{R11609118}
{3GPP Technical Specifications Group RAN}: {R1-1609118: RACH design with and
  without beam reciprocity} (Oct 2016)

\bibitem{R11609117}
{3GPP Technical Specifications Group RAN}: {R1-R11609117: Discussion on RA
  procedure} (Oct 2016)

\bibitem{TS38802}
{3GPP Technical Specifications Group RAN}: {TS 38.802 (Release 14): Physical
  Layer Aspects} (Sept 2017)

\bibitem{3GPPFeatures}
{3GPP Technical Specifications Groups}: {3GPP Features and Study Items} (2018),
  \url{http://www.3gpp.org/DynaReport/FeatureListFrameSet.htm}, accessed in
  June 2018

\bibitem{release15}
{3GPP Technical Specifications Groups}: {About Release 15} (June 2018),
  \url{http://www.3gpp.org/release-15}, accessed in June 2018

\bibitem{alkhateeb2014mimo}
Alkhateeb, A., Mo, J., Gonzalez-Prelcic, N., Heath, R.W.: {MIMO} precoding and
  combining solutions for millimeter-wave systems. IEEE Communications Magazine
   \textbf{52}(12),  122--131 (2014)

\bibitem{desai2014initial}
Desai, V., Krzymien, L., Sartori, P., Xiao, W., Soong, A., Alkhateeb, A.:
  {Initial beamforming for mmWave communications}. In: Signals, Systems and
  Computers, 2014 48th Asilomar Conference on. pp. 1926--1930 (2014)

\bibitem{RP161266}
{Deutsche Telekom AG}: {RP-161266: 5G architecture options -- full set} (June
  2016)

\bibitem{giordani2016initial}
Giordani, M., Mezzavilla, M., Zorzi, M.: {Initial access in 5G mmWave cellular
  networks}. IEEE Communications Magazine  \textbf{54}(11),  40--47 (2016)

\bibitem{ieee80211ay}
{IEEE P802.11 Task Group ay}: {Status of Project IEEE 802.11ay},
  \url{www.ieee802.org/11/Reports/tgay_update.htm}, accessed in June 2018

\bibitem{80211ad}
{IEEE Standard}: {802.11ad-2012: Enhancements for Very High Throughput in the
  60 GHz Band}, \url{https://ieeexplore.ieee.org/document/6392842/}

\bibitem{keysight2017understanding}
{Keysight Technologies}: {Understanding the 5G NR Rhysical Layer} (Nov 2017)

\bibitem{docomo2018}
{NTT Docomo}: {Status of Investigations on Physical-layer Elemental
  Technologies and High-frequency-band Utilization}. NTT Docomo Technical
  Journal  \textbf{19}(3) (2018)

\bibitem{Qualcomm2018Designing}
{Qualcomm}: {Designing 5G NR: The 3GPP Release-15 global standard for a
  unified, more capable 5G air interface} (Apr 2018),
  \url{https://www.qualcomm.com/media/documents/files/the-3gpp-release-15-5g-nr-design.pdf}

\bibitem{Qualcomm2018making}
{Qualcomm}: {Making 5G NR a reality}. White paper (Dec 2016)

\bibitem{razavizadeh2014three}
Razavizadeh, S.M., Ahn, M., Lee, I.: {Three-dimensional beamforming: A new
  enabling technology for 5G wireless networks}. IEEE Signal Processing
  Magazine  \textbf{31}(6),  94--101 (2014)

\bibitem{rohdeshwarz2017demystifying}
{Rohde \& Schwarz}: {Demystifying 5G -- How mobile is 5G at mmWave
  frequencies?} (2017),
  \url{https://www.rohde-schwarz.com/ru/solutions/test-and-measurement/wireless-communication/5g/webinars-videos/demystifying-5g-enable-mobility-in-5g-systems_231144.html},
  accessed in June 2018

\bibitem{rohdeshwarzslides}
{Rohde \& Schwarz}: {Numerology and Initial Access Concept for 5G NR} (2017),
  \url{https://www.youtube.com/watch?v=eE_b7vWbkoI}, accessed in June 2018

\bibitem{Yifei2017}
Yifei, Y., Xinhui, W.: {5G New Radio: Physical Layer Overview}. ZTE
  Communications  \textbf{15}(S1) (2017)

\end{thebibliography}
\vspace{-7px}

\end{document}